\documentclass[twocolumn,floats, prb, amsmath,showpacs]{revtex4}

\usepackage{epsfig}
\usepackage{colordvi}
\usepackage{graphicx}
\usepackage{bm}
\def\k{{\bf k}}

\def\R{{\bf R}}
\def\r{{\bf r}}
\def\A{{\bf A}}
\def\l{{\bf l}}
\def\K{{\bf K}}
\def\G{{\bf G}}
\def\O{{\mathbf \Omega}}
\def\a{{\bf a}}
\def\b{{\bf b}}
\def\la{\langle\kern-2.0pt\langle}
\def\ra{\rangle\kern-2.0pt\rangle}
\def\vt{\vert\kern-1.0pt\vert}

\begin{document}
\def\dvm#1{\marginpar{\small DV: #1}} \def\xwm#1{\marginpar{\small XW: #1}}
\def\ivo#1{\marginpar{\small Ivo: #1}}
\def\jym#1{\marginpar{\small JY: #1}}

\title{Fermi-surface calculation of the anomalous Hall conductivity}

\author{Xinjie Wang,$^1$ David Vanderbilt,$^1$ Jonathan R. Yates,$^{2,3}$
and Ivo Souza,$^{2,3}$}
\affiliation{$^1$Department of Physics and Astronomy, Rutgers University,
	Piscataway, NJ 08854-8019}
\affiliation{$^2$Department of Physics, University of California, Berkeley, CA 94720}	
\affiliation{$^3$Materials Science Division, Lawrence Berkeley National Laboratory, Berkeley, CA 94720}
\date{\today}
\begin{abstract}
While the intrinsic anomalous Hall conductivity is normally written
in terms of an integral of the electronic Berry curvature
over the occupied portions of the Brillouin zone, Haldane has
recently pointed out that this quantity (or more precisely,
its ``non-quantized part'') may alternatively be expressed as
a Fermi-surface property.  Here we present an {\it ab-initio}
approach for computing the anomalous Hall conductivity that takes
advantage of this observation by converting the integral over the
Fermi sea into a more efficient integral on the Fermi surface only.
First, a conventional electronic-structure calculation is performed
with spin-orbit interaction included. Maximally-localized Wannier
functions are then constructed by a post-processing step in order
to convert the {\it ab-initio} electronic structure around the
Fermi level into a tight-binding-like form. Working in the Wannier
representation, the Brillouin zone is sampled on a large number
of equally spaced parallel slices oriented normal to the total
magnetization. On each slice, we find the intersections of the
Fermi-surface sheets with the slice by standard contour methods,
organize these into a set of closed loops, and compute the
Berry phases of the Bloch states as they are transported around
these loops.  The anomalous Hall conductivity is proportional
to the sum of the Berry phases of all the loops on all the
slices. Illustrative calculations are performed for Fe, Co and Ni.

\end{abstract}
\pacs{71.15.Dx, 71.70.Ej, 71.18.+y, 71.20.Be, 75.47.-m.}
\maketitle

\vskip2pc
\marginparwidth 3.1in
\marginparsep 0.5in


\section{Introduction}

It is by now well established that the intrinsic Karplus-Luttinger 
mechanism\cite{karplus54}
plays a significant role in the anomalous Hall conductivity (AHC)
of ferromagnets. This contribution can be expressed as an integral
of the $k$-space Berry curvature over the occupied portions of the
Brillouin zone (BZ).\cite{chang96,sundaram99,onoda02,jungwirth02}
First-principles calculations of the intrinsic AHC have been
carried out by several authors, using either a Kubo linear-response
formula\cite{fang03,yao04} or a direct ``geometric'' evaluation
of the Berry curvature,\cite{wang06}
and achieving good agreement with experimental values for several
ferromagnets. These studies revealed that the Berry curvature is
very sharply peaked in certain regions of the BZ where
spin-orbit splitting occurs near the Fermi level.  As a result
the calculations tend to be rather demanding; in the case of
bcc Fe, for example, millions of $k$-points must be sampled to
achieve convergence.\cite{yao04}   More efficient approaches are
therefore highly desirable.

In a preceding paper,\cite{wang06} we developed a strategy for
calculating the AHC in which Wannier interpolation of the Bloch
functions was used to circumvent the need to perform a full
first-principles calculation for every $k$-point.  Thus, while
the required number of $k$-points was not reduced, the computational
load per $k$-point was greatly reduced.
In this approach, the actual first-principles calculations
are performed on a comparatively coarse $k$-mesh. Then, in a
postprocessing step, the calculated electronic structure is mapped
onto an ``exact tight-binding model'' based on maximally-localized
Wannier functions.\cite{souza01} Working in the Wannier representation, the
Berry curvature can then be evaluated very inexpensively at each
of the $k$-points of the fine mesh needed for accurate evaluation
of the AHC.

Recently, Haldane has shown that while the intrinsic AHC is usually
regarded as a Fermi-sea property of all the occupied states, it
can alternatively, and in some ways more naturally, be regarded
as a Fermi-surface property.\cite{haldane04}  (More precisely,
Haldane showed that these quantities are equal modulo the quantum
of transverse conductivity that is well-known from the quantum
Hall effect, since one cannot rule out the possibility that, e.g.,
some occupied bands carry non-zero Chern numbers.\cite{haldane04})
By a kind of integration by parts, Haldane showed how the
integral of the Berry curvature over the occupied portions of the BZ
could be manipulated first into a Fermi-surface integral of a
Berry connection, and then ultimately into a Fermi-surface integral
of a Fermi-vector-weighted Berry curvature, augmented with some
Berry-phase corrections for the case of non-simply-connected Fermi sheets.
In discussing his Eq.~(23), Haldane mentions in passing that
this expression can also be reformulated in terms of the
Berry phases of electron orbits circulating on the Fermi surface.

In this paper we present a tractable and efficient computational scheme
that takes the latter point of view as its organizing principle.
In our approach, the BZ is divided into a fine mesh
of equally-spaced slices normal to the direction of the magnetization,
and the integral of the Berry curvature
over the occupied states of a given slice is transformed into a
sum of Berry phases of Fermi loops lying in that slice.  As a result, 
the three-dimensional BZ integration is avoided, and the method
relies instead only on information calculated on the two-dimensional
Fermi surface.  As in Ref.~\onlinecite{wang06}, an important
ingredient of our approach is the use of a Wannier interpolation
scheme to lower the cost further by eliminating the need for a full
first-principles evaluation at each point on the Fermi surface.
Combining these two complementary strategies, we arrive at a
robust and efficient method for computing of the AHC
in ferromagnetic metals.

The paper is organized as follows. In Sec.~\ref{sec:method} we
present the necessary formulas relating Berry phases on the Fermi surface to 
the AHC, as well as their evaluation in the Wannier representation. 
The details of the first-principles calculations and the determination
of the Fermi loops are given in Sec.~\ref{sec:comp}. In
Sec.~\ref{sec:results} the method is applied to the
transition metals Fe, Co and Ni.
A discussion of issues of computational efficiency is given
in Sec.~\ref{sec:disc},
followed by a brief conclusion in Sec.~\ref{sec:summary}.

\section{Method}
\label{sec:method}

\subsection{Fermi-loop formula}
\label{sec:floop}

Our starting point is the AHC expressed as an
antisymmetric Cartesian tensor in terms of the Berry curvature,
\begin{eqnarray}
\sigma_{\alpha \beta}=-\frac{e^2}{\hbar}\sum_{n}\int_{\rm BZ}\,
\frac{d\k}{(2\pi)^3} \, f_n(\k)\,\Omega_{n,\alpha \beta}(\k) \;,
\label{eq:sigma}
\end{eqnarray}
where the integration is over the three-dimensional 
BZ and the occupation function $f_n(\k)$ restricts the sum
to the occupied states (we work at zero temperature).
$\Omega_{n,\alpha \beta}(\k)$ is the Berry-curvature matrix of 
band $n$, defined as
\begin{eqnarray}
	\Omega_{n,\alpha \beta}(\k) =
        -2\,{\rm Im}\,\Big\langle \frac{\partial u_{n\k}}
        {\partial k_\alpha}
        \Big|\frac{\partial u_{n\k}}{\partial k_\beta} \Big\rangle
\end{eqnarray}
where $u_{n\k}$ is the periodic part of the Bloch function $\psi_{n\k}$.
Because $\Omega_{n,\alpha \beta}$ is antisymmetric, we
can represent it instead in axial-vector notation as
\begin{equation}
\Omega_{n\gamma}= \frac{1}{2} \sum_{\alpha\beta} \epsilon_{\alpha\beta\gamma}
\Omega_{n,\alpha \beta} \;,
\end{equation}
or equivalently,
$\Omega_{n,\alpha \beta}= \sum_\gamma \epsilon_{\alpha\beta\gamma}
\Omega_{n\gamma}$, where $\epsilon_{\alpha\beta\gamma}$ is
the antisymmetric tensor.  The Berry curvature can also be written as
\begin{equation}
\O_n(\k)=\nabla_\k\times\A_n(\k)
\label{eq:curl}
\end{equation}
where the Berry connection is
\begin{equation}
{\bf A}_{n}(\k)=i\langle u_{n\k}|\nabla_\k|u_{n\k} \rangle\;.
\label{eq:berry_connection}
\end{equation}

Following Ref.~\onlinecite{haldane04}, we rewrite Eq.~(\ref{eq:sigma}) as
\begin{equation}
\sigma_{\alpha \beta}=\frac{-e^2}{\hbar}\frac{1}{(2\pi)^2}
\sum_{n\gamma}^{} \epsilon_{\alpha\beta\gamma} K_{n\gamma}
\label{eq:sigmaK}
\end{equation}
where
\begin{equation}
\K_n = \frac{1}{2\pi}\int_{\rm BZ}\, d\k \, f_n(\k)\,\O_n(\k) \;.
\label{eq:Kdef}
\end{equation}
For the case of a completely filled band lying entirely below the
Fermi level, Haldane has shown\cite{haldane04} that ${\bf K}_n$ is
quantized to be a reciprocal lattice vector (the ``Chern vector''), as
will become clear in Sec.~\ref{sec:quantum} below.

Let $\a_i$ and $\b_i$ be a conjugate set of primitive
real-space and reciprocal-space lattice vectors respectively,
$\a_i\cdot\b_j=2\pi\delta_{ij}$, and let
\begin{equation}
c_{nj}=\frac{1}{2\pi}\,\a_j\cdot\K_n
\label{eq:cnjdef}
\end{equation}
so that
\begin{equation}
\K_n = \sum_j c_{nj}\,\b_j \;.
\label{eq:Kn}
\end{equation}
In order to compute
$c_{n3}$, for example, we choose the BZ to be a prism whose base is
spanned by $\b_1$ and $\b_2$ and whose height is $2\pi/a_3$, and
convert the integral into one over slices parallel to the base.
In general, separate calculations in which the slices are
constructed parallel to the $\b_2$-$\b_3$ and $\b_1$-$\b_3$
planes are needed to compute $c_{n1}$ and $c_{n2}$
respectively.\cite{explan-pol}
However, this can be avoided in the
common case that the magnetization lies parallel to a symmetry
axis; one can then choose $\b_1$ and $\b_2$ perpendicular
to this axis, and only $c_{n3}$ needs to be computed.

Inserting Eq.~(\ref{eq:Kdef}) into Eq.~(\ref{eq:cnjdef}) yields
\begin{equation}
c_{nj}= \frac{a_j}{2\pi}\int_0^\frac{2\pi}{a_j} dk_\perp
   \frac{\phi_n(k_\perp)}{2\pi}
\label{eq:Cshort}
\end{equation}
where
\begin{equation}
\phi_n(k_\perp)= 
   \int_{{\cal S}_n(k_\perp)} d^2k \; \hat{a}_j\cdot\O_n(\k) \;.
\label{eq:phibc}
\end{equation} 
Here $k_\perp$ labels the slice
and ${\cal S}_n(k_\perp)$ is the region of the slice
in which band $n$ is occupied.  Recalling Eq.~(\ref{eq:curl})
and noting that $\hat{a}_j$ is the unit vector normal to the slice, the
application of Stokes' theorem to Eq.~(\ref{eq:phibc}) yields
\begin{equation}
\phi_n(k_\perp) = \oint_{{\cal C}_n(k_\perp)} \A_n(\k)\cdot d{\bf l}
\label{eq:philoop}
\end{equation}
where ${\cal C}_n(k_\perp)$ is the oriented curve bounding
${\cal S}_n(k_\perp)$ on the slice and $\phi_n(k_\perp)$ has
the interpretation of a Berry phase.  For later convenience we
also define
\begin{equation}
\phi(k_\perp)=\sum_n \phi_n(k_\perp)
\label{eq:phitot}
\end{equation}
and similarly $c_j=\sum_n c_{nj}$ etc.
The calculation of the AHC has thus been reduced to a calculation
that is restricted to the Fermi surface only, in the spirit of
Eq.~(23) of Haldane's Ref.~\onlinecite{haldane04}.

\begin{figure}
\begin{center}
   \epsfig{file=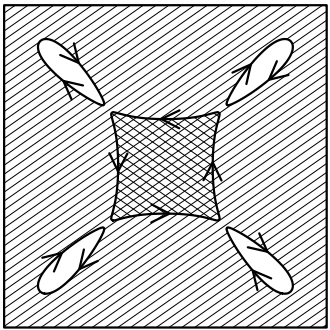,width=1.7in}
\end{center}
\caption{Sketch of intersections of the Fermi surface with
a constant-$k_\perp$ plane.  Open, hashed, and
cross-hashed regions correspond to filling of zero, one, and two
bands, respectively.  The four small Fermi loops belong to the first band,
while the large central one belongs to the second.  Arrows indicate
sense of circulation for performing the Berry-phase integration.}
\label{fig:loop_demo}
\end{figure}

In general, the occupied or unoccupied region of band $n$ in
slice $k_\perp$ need not be simply connected, in which case the
boundary ${\cal C}_n(k_\perp)$ is really the union of several
loops.  Moreover, loops encircling hole pockets should be taken
in the negative direction of circulation.  This is illustrated
in Fig.~\ref{fig:loop_demo}, where the first band exhibits four hole pockets
and the second band has one electron pocket, so that ${\cal C}_1$ is the union
of four countercirculating loops and ${\cal C}_2$ is a fifth
loop of positive circulation.  If higher bands are
unoccupied, then $\phi(k_\perp)$ for this slice is
just given by the sum of the Berry phases of these five loops.
We shall assume for simplicity in the following that
${\cal C}_n(k_\perp)$ is simply connected, but the generalization
to composite loops is straightforward.

\subsection{The quantum of Hall conductivity}
\label{sec:quantum}

We claimed earlier that if band $n$ is fully occupied, $\K_n$
in Eq.~(\ref{eq:Kdef}) is quantized to a reciprocal lattice
vector.  This can now be seen by noting that 
under those circumstances the integral in
Eq.~(\ref{eq:phibc}) runs over a two-dimensional BZ, which can be
regarded as a closed two-dimensional manifold (two-torus), and for
topological reasons\cite{tknn} the integral of the Berry curvature
over such a closed manifold must be an integer multiple of $2\pi$
(the Chern number).  Then each $c_{nj}$ is an integer, and $\K_n$
in Eq.~(\ref{eq:Kn}) must be a reciprocal lattice vector as
claimed.  If the system is an insulator, then $\K=\sum_n\K_n$
(summed over occupied bands) is also guaranteed to be a reciprocal
lattice vector, and if it is a nonzero one, the insulator would
have a quantized Hall conductivity and could be regarded as a
quantum Hall crystal (or ``Chern insulator'').\cite{haldane04,haldane88}
No physical realization of such a system is known experimentally,
but the search for one remains an interesting challenge.

Let us consider again a slice for which band $n$ is fully occupied
but has a non-zero Chern number.  If this slice is regarded as an
open rectangle (or parallelogram) rather than a closed two-torus,
and a continuous choice of gauge is made in its interior (i.e.,
$\A_n(\k)$ is free of singularities), then the boundary ${\cal
C}_n(k_\perp)$ is the perimeter of this rectangle and
Eq.~(\ref{eq:philoop}) will yield the same integer multiple of
$2\pi$ as Eq.~(\ref{eq:phibc}).  In the spirit of
Fig.~\ref{fig:loop_demo}, however, we prefer to regard the slice as
a closed two-torus and to exclude the perimeter from our
definition of the boundary ${\cal C}_n(k_\perp)$.  Then ${\cal
C}_n(k_\perp)$ is null and Eq.~(\ref{eq:philoop}) vanishes for the
case at hand, in disagreement with Eq.~(\ref{eq:phibc}).  The
disagreement arises because of the impossibility of making a
continuous choice of gauge on a closed manifold having a non-zero
Chern number;\cite{tknn} the best that can be done is to make $\A_n(\k)$
finite everywhere except at singularities (``vortices'') which,
when included, restore the missing contributions of
$2\pi$.

Returning to the general case of a partially occupied band $n$
with ${\cal C}_n(k_\perp)$ defined to exclude the perimeter of the
slice, we conclude that Eq.~(\ref{eq:philoop}) is
really only guaranteed to equal the true result of Eq.~(\ref{eq:phibc})
modulo $2\pi$.  Moreover, the Berry phase will be
evaluated in practice using
a discretized Berry-phase formula\cite{ksv93} of the form
\begin{equation}
\phi_n(k_\perp) = -{\rm Im}\,{\ln}\,\prod_j
 \langle u_{n\k_j} \vert u_{n\k_{j+1}}\rangle
\label{eq:phibp}
\end{equation}
where $\k_j$ discretizes the loop ${\cal C}_n(k_\perp)$.  (We will
actually use a modified version, Eq.~(\ref{eq:bp}), of this
formula.)  The choice of branch cut is now arbitrary, and again the
agreement with Eq.~(\ref{eq:philoop}) or Eq.~(\ref{eq:phibc}) is
only guaranteed modulo $2\pi$.  By convention one normally
restricts phases to lie in the interval $(-\pi,\pi]$, but then
$\phi_n(k_\perp)$ would in general have unwanted discontinuities
at some values of $k_\perp$.  In practice we discretize the
$k_\perp$ integration, so that using Eq.~(\ref{eq:Cshort}),
$c_j=\sum_n c_{nj}$ becomes
\begin{equation}
c_j= \frac{1}{n_{\rm slice}} \sum_{i=1}^{n_{\rm slice}}
   \frac{\phi(i)}{2\pi} \;.
\label{eq:csum}
\end{equation}
We then enforce continuity of the total phase $\phi(k_\perp)$ of
Eq.~(\ref{eq:phitot}) by choosing $\phi(i)$ such that
$|\phi(i)-\phi(i-1)|\ll 2\pi$ for each slice $i=2,3,...$ in
sequence.  Since the true phase given by the sum of contributions
in Eq.~(\ref{eq:phibc}) is also continuous, this guarantees that
our calculated $\phi(k_\perp)$ differs from the true one by the
same multiple of $2\pi$ for all $k_\perp$.  Our computed
AHC would then differ from the true one by a multiple of the
quantum and could be said to give the ``non-quantized part''
of the intrinsic AHC in the sense of Haldane.\cite{haldane04}
However, it is straightforward to remove this overall ambiguity of branch
choice by evaluating $\phi(k_\perp)$ from Eq.~(\ref{eq:phibc}) on
the first slice and then enforcing continuity for each
subsequent slice, thus arriving
at the correct AHC without any question of a quantum.

We note in passing that
an isolated point of degeneracy (``Dirac point'')
between a pair of bands $n$ and $n+1$ can generically occur
in three-dimensional $k$-space in the absence of
time-reversal symmetry.\cite{haldane04} If such a Dirac point occurs below
the Fermi energy, then $\phi_n(k_\perp)$ and $\phi_{n+1}(k_\perp)$
will, when evaluated from Eq.~(\ref{eq:phibc}),
exhibit equal and opposite discontinuities of $2\pi$ at the
$k_\perp$ of the Dirac point.  However, the total phase $\phi(k_\perp)$
will remain continuous, so that the algorithm described in the
previous paragraph will still work correctly.

We close this subsection by emphasizing that the discussion of
possible non-zero Chern numbers or the presence of Dirac points is
rather academic.  In our calculations on Fe, Ni and Co, we have not
encountered any indications of such anomalies; they presumably
occur rarely or not at all in the materials studied here.

\subsection{Evaluation of the Fermi-loop Berry phase}

The essential problem now becomes the computation of the loop integral of
Eq.~(\ref{eq:philoop}).  As is well known, the Berry connection
$\A_n(\k)$ of Eq.~(\ref{eq:berry_connection}) is gauge-dependent,
i.e., sensitive to 
the $k$-dependent choice of phase of the Bloch
functions.  If  Eq.~(\ref{eq:philoop}) is to be calculated by the
direct evaluation of $\A_n(\k)$ and its subsequent integration
around the loop, this lack of gauge-invariance may present
difficulties. For example, it means that there is no unique
Kubo-formula expression for $\A_n(\k)$.  An alternative and more
promising approach is to compute $\phi_n(k_\perp)$ by the discretized
Berry-phase formula\cite{ksv93} of Eq.~(\ref{eq:phibp}), where the
inner products are computed from the full first-principles
calculations at neighboring pairs of $k$-points around
the loop.  However, this may still be quite time-consuming if it
has to be done at very many $k$-points.  We avoid this by
using the technique of Wannier interpolation\cite{souza01,wang06,yates07}
to perform the needed loop integral inexpensively.  In this formulation,
the loop integral of Eq.~(\ref{eq:philoop}) can be expressed as
a sum of two terms, one in which a contribution to $\A_n(\k)$
is evaluated and integrated explicitly, and a second that takes a form like that 
of Eq.~(\ref{eq:phibp}).

The key idea of Wannier interpolation is to map the low-energy
first-principles electronic structure onto an ``exact tight-binding
model'' using a basis of appropriately constructed crystalline
Wannier functions.  For metallic systems like those
considered here, the bands generated by these Wannier functions are
only partially occupied.
They are guaranteed by construction to reproduce the true first-principles 
bands in an energy
window extending somewhat above the Fermi level, so that all valence and 
Fermi-surface states are properly
described.\cite{souza01} In the Wannier representation, the
desired quantities such as band energies, eigenstates and the
derivatives of eigenstates with respect to wavevector $k$ can then
be evaluated at arbitrary $k$-points at very low computational
cost. All that is needed is to evaluate, once and for all, the
Wannier-basis matrix elements of the Hamiltonian and position
operators.\cite{wang06} 
It is worth pointing out that it may sometimes be expedient
to drop some lower occupied bands and construct the Wannier functions
so that they correctly represent the Bloch functions only in some
narrower energy window containing the Fermi energy; since the
present formulation involves only Fermi-surface properties, the
nonquantized part of the AHC will then still be given correctly.

The Wannier construction procedure of Ref.~\onlinecite{souza01}
provides us with a set of $M$ Wannier functions $|\R n\rangle$
($n=1,...,M$) in each cell labeled by lattice vector $\R$.  From
these the Bloch basis functions $|u_{n\k}^{\rm(W)}\rangle$ are
constructed according to the Fourier transform relation
\begin{eqnarray}
|u_{n\k}^{\rm(W)}\rangle=\sum_{\R} e^{-i\k \cdot(\r-\R)}|\R n\rangle\ .
\label{eq:ft}
\end{eqnarray}
Here the superscript $(\rm W)$ indicates that these are obtained from
the Wannier representation,  that is, they are not yet Hamiltonian
eigenstates.  To obtain those, we construct the $M\times M$ Hamiltonian
matrix
\begin{eqnarray}
H^{\rm(W)}_{nm}(\k)=\langle u_{n\k}^{\rm(W)}|\hat H(\k)|u_{m\k}^{\rm(W)}\rangle\
\end{eqnarray}
via
\begin{equation}
H_{nm}^{\rm(W)}=\sum_\R e^{i\k\cdot\R}\;\langle{\bf 0}n|\hat{H}|\R m\rangle
\,.
\label{eq:uu}
\end{equation}
At any given $\k$ this matrix can be diagonalized to yield an $M\times M$
unitary matrix $U_{nm}(\k)$, i.e.,
\begin{eqnarray}
U^{\dagger}(\k)H^{\rm(W)}(\k)U(\k)=H^{\rm(H)}(\k)
\end{eqnarray}
where $H^{\rm(H)}(\k)={\cal E}^{\rm(H)}_n \delta_{mn}$ are the energy
eigenvalues and
\begin{eqnarray}
| u_{n\k}^{\rm(H)}\rangle= \sum_{m}| u_{m\k}^{\rm (W)} \rangle U_{mn}(\k)
\label{eq:uwu}
\end{eqnarray}
are the corresponding band states.  By the construction procedure
of Ref.~\onlinecite{souza01}, ${\cal E}^{\rm(H)}_n$ is identical
to the true ${\cal E}_n$ (and similarly for the eigenvectors
$u_{n\k}^{\rm(H)}$) for all occupied states and low-lying empty states.
This is strictly true only for $k$-points on the original {\it ab-initio}
mesh. The power of this interpolation scheme lies in the fact that, by virtue 
of the spatial localization of the Wannier functions,
the error remains extremely small even for points away from that
grid.\cite{yates07}

The next step is to evaluate ${\cal E}^{\rm(H)}_{n\k}$
on a two-dimensional mesh of $k$-points covering a single slice
and then
use a contour-finding algorithm to map out and discretize the Fermi
loops therein.  This part of our scheme will be
described in more detail in Sec.~\ref{sec:contour}.  For now we can
just assume that the output is a sequence of points $\k_j$
$(j=0,\ldots,J-1)$ 
providing a fairly dense mapping of the contour.
(As before, we assume for simplicity that the Fermi
contour consists of a single loop; the extension to multiple loops
is straightforward.)

Next we need to obtain the Berry connection
$\A_{n}(\k)=i\langle u_{n\k}^{\rm(H)}|\nabla_\k|u_{n\k}^{\rm(H)} \rangle$
as in
Eq.~(\ref{eq:berry_connection}).  Using Eq.~(\ref{eq:uwu}), this
becomes
\begin{eqnarray}
\nonumber
\A_n(\k)&=& \sum_{lm}U^{\dagger}_{nl}(\k) \, \A_{lm}^{\rm(W)}(\k)
	 U_{mn}(\k)\\
	&&+i\sum_m U^{\dagger}_{nm}(\k)\nabla_{\k}U_{mn}(\k)
\label{eq:Adecomp}
\end{eqnarray} 
where
\begin{equation}
\A_{nm}^{\rm(W)}(\k)=
	i\langle u_{n\k}^{\rm(W)}|\nabla_\k|u_{m\k}^{\rm(W)}\rangle
\end{equation}
is computed in practice from the expression
\begin{equation}
\A_{nm}^{\rm(W)}(\k)=\sum_\R e^{i\k\cdot\R}\;
   \langle{\bf 0}n|\hat{\bf r}|\R m\rangle
\label{eq:ww}
\end{equation}
in a manner similar to Eq.~(\ref{eq:uu}).  Details concerning the method
of calculating Eqs.~(\ref{eq:uu}) and (\ref{eq:ww}) can be found in
Ref.~\onlinecite{wang06}.

The decomposition of $\A_n(\k)$ into two terms in Eq.~(\ref{eq:Adecomp})
is an artifact of the choice of Wannier functions; only the sum of the
two terms is physically meaningful (upon a circuit 
integration).  However, for a given choice
of Wannier functions, the first term
arises because the Bloch functions $|u_{n\k}^{\rm(H)}\rangle$
acquire some of the Berry curvature attached to the full subspace
of $M$ Wannier functions used to represent them, whereas the
second term represents the Berry curvature arising from changes
of character of this Bloch state {\it within} the Wannier subspace.
To clarify this viewpoint, we introduce a notation\cite{wang06} in which
$\vt v_{n\k}\ra$ is defined to be the $n$th column vector of matrix
$U$, so that the second term of Eq.~(\ref{eq:Adecomp}) becomes
$i\la v_{n\k}\vt\nabla_\k\vt v_{n\k}\ra$.  Plugging into
Eq.~(\ref{eq:philoop}), this yields
\begin{eqnarray}
\phi_n(i)&=&\oint \la v_{n\k} \vt{\bf A}^{\rm(W)}(\k)\vt v_{n\k}\ra
 \cdot d\l
\nonumber\\&&
+i\oint \la v_{n\k} \vt\nabla_\k\vt v_{n\k}\ra\cdot d\l
\label{eq:bpcontin}
\end{eqnarray}
for the Berry phase of slice $i$ appearing in Eq.~(\ref{eq:csum}).
Note that the integrand in the first term is gauge-invariant
(here ``gauge'' refers to the application of a phase twist
$\vt v_{n\k}\ra\rightarrow e^{i\beta(\k)}\,\vt v_{n\k}\ra$), while
in the second term only the entire loop integral is
gauge-invariant.  Indeed, the
second term is just a Berry phase defined within the
$M$-dimensional ``tight-binding space'' provided by the
Wannier functions.  Recalling that $\k_j$ for $j=0,\ldots,J-1$
is our discretized description of the Fermi loop,
and using standard methods for discretizing Berry phases\cite{ksv93} as
in Eq.~(\ref{eq:phibp}), our final result becomes
\begin{eqnarray}
\phi_n(i)&=&\sum_{j=0}^{J-1} \,
\la v_{n\k} \vt {\bf A}^{\rm(W)}(\k)\vt v_{n\k}\ra \cdot \Delta\k
\nonumber\\&&
-{\rm Im}\,{\ln}\,\prod_{j=0}^{J-1}\la v_{n\k_j} \vt v_{n\k_{j+1}}\ra,
\label{eq:bp}
\end{eqnarray}
where $\Delta\k=(\k_{j+1}-\k_{j-1})/2$.

As we shall see below, in practice we only encounter closed orbits,
in which case it is clearly appropriate to set $\k_J=\k_0$ and close
the phases with $\vt v_{n,\k_J}\ra=\vt v_{n,\k_0}\ra$.
For lower-symmetry situations, however, open orbits with
$\k_J=\k_0+\G$ may be encountered.  Even in this case,
however, we would still set $\vt v_{n,\k_J}\ra=\vt
v_{n,\k_0}\ra$; in contrast to the full Bloch states which obey\cite{ksv93}
$u_{n,\k_J}=e^{-i\G\cdot\r}\,u_{n,\k_0}$, no extra phase factors
are needed here because the Fourier-transform convention of
Eq.~(\ref{eq:ft}) treats the Wannier functions as though they are
all nominally located at the cell origin.

In summary, our strategy is to evaluate Eq.~(\ref{eq:csum}) by
decomposing each generalized path ${\cal C}_n(i)$ into connected
simple loops, and sum the loop integrals as computed using
Eq.~(\ref{eq:bp}).  The operations needed to evaluate Eq.~(\ref{eq:bp})
are inexpensive as they all involve vectors and matrices defined
in the low-dimensional space of the Wannier representation.

\section{Computational details}
\label{sec:comp}

\subsection{First-principles calculations}

Fully relativistic band-structure calculations are carried out
for the ferromagnetic transition metals Fe, Co and Ni at their
experimental lattice constants (5.42, 4.73, and 6.65 bohr,
respectively) using the {\tt PWSCF}
code.\cite{pwscf} Norm-conserving pseudopotentials with 
spin-orbit coupling\cite{corso05} are generated using similar parameters as 
in Ref.~\onlinecite{wang06}. An energy cutoff of 60 Hartree is used for 
the planewave expansion of the valence wavefunctions 
(400 Hartree for the charge densities), and the PBE 
generalized-gradient approximation\cite{perdew96} 
is used for the exchange-correlation functional. 
The self-consistent ground state is obtained using a $16\times
16\times 16$ Monkhorst-Pack\cite{mp} mesh of $k$-points and a fictitious
Fermi smearing\cite{coldsmear} of $0.02$ Ry for the 
Brillouin-zone integration.

\begin{table}
\caption{Calculated spin magnetic moment per atom (in $\mu_{\rm B}$)
for the three transition metals Fe, Ni and Co, with magnetization
along [001], [111] and [001], respectively.}
\begin{ruledtabular}
\begin{tabular}{lccc}
& bcc Fe & fcc Ni & hcp Co \cr 
\hline
Theory & 2.22 & 0.62 & 1.60 \cr
Experiment\footnotemark[1]  & 2.13 & 0.56 & 1.59
\end{tabular}
\end{ruledtabular}
\footnotetext[1]{Ref.~\onlinecite{daalderop90}.}
\label{table:spin_moment}
\end{table}

The calculated spin magnetic moments are shown in
Table~\ref{table:spin_moment}.  The effect of spin-orbit
coupling on these moments is included in the calculation, since
it is needed in any case to obtain a nonzero AHC.
The agreement with experiment is rather good, confirming that
our norm-conserving pseudopotentials are suitable for describing
the ferromagnetic state of the transition metals.

The maximally-localized Wannier functions are generated using 
the {\tt WANNIER90} code;\cite{wannier90} details are given
in Secs.~\ref{sec:fe}-\ref{sec:co} below.

\subsection{Mapping and sampling of Fermi loops}
\label{sec:contour}

As discussed above, our basic strategy involves dividing the BZ
into a series of parallel slices and finding the intersections
of the Fermi surface with each of these slices.  Each slice is
sampled on a uniform $N\times N$ $k$-point mesh, with $N$ ranging
from $300$ to $500$, and the band energies are computed on the
mesh using Wannier interpolation.  A standard contour-finding
algorithm of the kind used to make contour plots is then used
to generate a list of Fermi loops and, for each loop, a list
$\k_0,\ldots,\k_{J-1}$ of $k$-points providing a discretized
representation of the loop.

\begin{figure}
\begin{center}
\epsfig{file=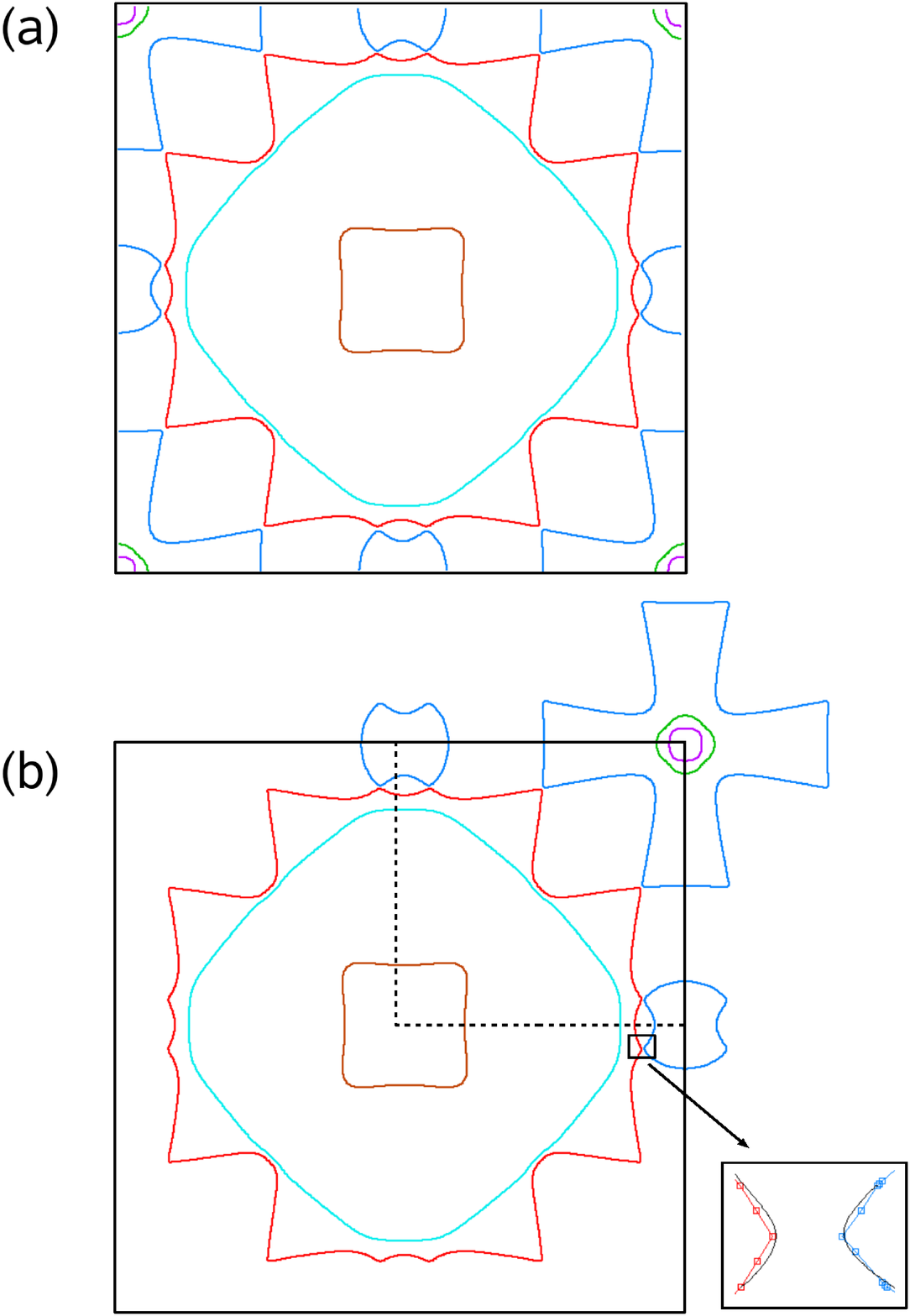,width=3.0in}
\end{center}
\caption{(Color online) Calculated Fermi-surface intersections
(Fermi loops) on the $k_z=0.02$ plane for bcc Fe; different bands
are color-coded for clarity.
(a) Fermi contours within the first Brillouin zone.
(b) Fermi contours after reassembly to form closed contours
by translating some portions by a reciprocal lattice vector.
Inset: enlargement showing part of an avoided crossing where
a refined mesh (black lines) is used to obtain 
a more accurate representation of the Fermi loop.
The actual calculation is performed within the dashed box.
}
\label{fig:slice}
\end{figure}

As shown in Fig.~\ref{fig:slice}(a), the Fermi contours in the
first BZ are sometimes composed of multiple segments terminating
at the BZ boundary.  To insure that we get closed loops suitable
for the evaluation of Eq.~(\ref{eq:bp}), we actually do the initial
contour-finding procedure in an extended zone with $3\times 3$ times
the size of the first BZ.  We then select closed loops located
near the central cell while identifying and discarding loops or
portions of loops that correspond to periodic images of these
chosen loops.  The result is a set of closed loops that partially
extend outside the first BZ as shown in Fig.~\ref{fig:slice}(b).
Of course, if there were open orbits on the Fermi surface, it would
not always be possible to select closed loops in the above sense;
one would have to accept a ``loop'' with $\k_J=\k_0+\G$
as discussed following Eq.~(\ref{eq:bp}).  However, we never
encounter such open orbits in practice for the types of materials
studied here, in which the magnetization is aligned with a three-fold,
four-fold, or six-fold rotational symmetry axis.  The slices are
perpendicular to the symmetry axis, and the symmetry ensures
that open orbits cannot occur on the slices.

A potential difficulty in applying the Fermi-loop method to
real materials arises from the possible presence of degeneracies or
near-degeneracies between bands.  If two bands are degenerate
at the Fermi energy, this means that two Fermi loops touch,
and it is no longer straightforward to define and compute the
Berry phases of these loops.  Fortunately, the presence of
ferromagnetic spin splitting and spin-orbit coupling removes
almost all degeneracies. In our calculations we found no true
degeneracies in hcp Co or fcc Ni, and the only degeneracies in bcc
Fe were found to lie in the $k_z=0$ plane.  (In the latter case,
we avoid the $k_z=0$ plane by picking a $k_\perp$ mesh that is offset
so that this plane is skipped over.) On the other hand, we
do find numerous weakly avoided crossings induced by the spin-orbit
interaction, and while these introduce no difficulty in principle,
they do require special care in practice.  Indeed, we find that
it is important to sample the Fermi surface very accurately in
the vicinity of these crossings.  To do so, we calculate the Berry
curvature at each $\k_j$ using Wannier interpolation, and
if a large value is encountered, we introduce a refined mesh
with $4\times 4$ greater density in this region, repeat the
contour-finding procedure there, and replace the discretized
representation of this portion of the loop with a denser one.
We also take care to recompute ${\cal E}_{n\k}$ at each $\k_j$
and iteratively adjust the $k$-point location in the
direction transverse to the loop
in order to insure that ${\cal E}_{n\k}$ lies precisely
at the Fermi energy.  An example of a portion of a Fermi loop
that has been refined in this way is illustrated in the inset
to Fig.~\ref{fig:slice}(b).  Overall, the resulting number $J$
of $k$-points per loop ranges from several hundreds to thousands,
depending on the size and complexity of the Fermi loop.

In our current implementation, the entire procedure above is
repeated independently on each of the slices.  As already mentioned
in Sec.~\ref{sec:quantum}, it is important to make a
consistent choice of branch of the Berry phase $\phi(i)$ on
consecutive slices.  We do this by adding or subtracting a multiple
of $2\pi$ to the Berry phase calculated from Eq.~(\ref{eq:csum}) such
that $|\phi_n(i)-\phi_n(i-1)|\ll2\pi$ is satisfied, always checking
for consistency between the first and last slice.

\subsection{Use of symmetry to reduce computational load}
\label{sec:sym}

The presence of a net magnetization results in a considerable
reduction in symmetry, but several symmetries still remain
that can be exploited to reduce the computational cost.  In the
previous Fermi-sea-based methods\cite{yao04,wang06} the use of
symmetries is straightforwardly implemented by restricting
the $k$-point sampling to the irreducible wedge of the BZ. For
the Fermi-loop method, the use of symmetries needs more careful
treatment.

Here we discuss the difficulties, and point out their solution,
using ferromagnetic bcc Fe as an example.  We focus our attention
on the mirror symmetries $M_x$ and $M_y$. Since each slice lies in
an $x$-$y$ plane, we can use these to restrict the
bandstructure calculation and the search for Fermi contours
to a reduced BZ having one-fourth of the area of the full BZ,
as shown by the dashed line in Fig.~\ref{fig:slice}(b).  However,
a typical Fermi loop will no longer close within this reduced BZ.
Because a Berry phase is a global property of a closed loop,
one cannot just compute the Berry phase of open segment lying inside
the reduced BZ and multiply by four; the Berry
phase of this segment is ill-defined unless the phases of the
wavefunctions at its terminal points are specified.

\begin{figure}
\begin{center}
\epsfig{file=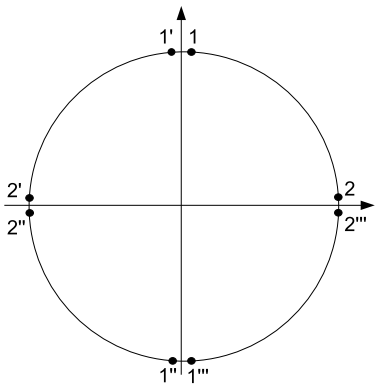,width=2.0in}
\end{center}
\caption{Illustration of use of $M_x$ and $M_y$ mirror symmetries
on a slice of the Brillouin zone in bcc Fe.  Only the segment of
the Fermi loop from Point 2 to Point 1 is actually computed; the
three other segments are included using symmetry operations.}
\label{fig:loop_sym}
\end{figure}

Our solution to this difficulty is illustrated in
Fig.~\ref{fig:loop_sym}.  We make some arbitrary but
definite choice of the phases of the Bloch functions in the
upper-right segment, compute the open-path Berry phase following
Eq.~(\ref{eq:bp}), and multiply by four.
We then add corrections that take account of the phase jumps at
the segment boundaries.  For example, we let $M_x$ acting on the
Bloch states from $1$ to $2$ define the Bloch states
from $1'$ to $2'$.  The correction
arising from the $1'$-$1$ boundary is then given by the phase of
$\langle u_{1'}\vert u_1\rangle=\langle M_x u_{1}\vert u_1\rangle$.
(Here $M_x$ is defined in the spinor context and includes a complex
conjugation component.
Since the Bloch functions are expressed in the Wannier basis
in our approach, information about the symmetries of the Wannier
functions has to be extracted and made available for the application
of the symmetry transformations.)
Similar corrections, using also $M_y$, are obtained for the
$2'$-$2''$, $1''$-$1'''$, and $2'''$-$2$ segment boundaries.
By including these mismatch
corrections, we are able to calculate the global Fermi-loop
Berry phase in a correct and globally gauge-invariant manner.

We have tested this procedure and confirmed that the results
obtained are essentially identical to those computed without
the use of symmetry.  The BZ could in principle be reduced
further in bcc Fe using the diagonal mirror operations, but we
have not tried to implement this.

\section{Results}
\label{sec:results}

In this section we present the results of our calculations of 
the anomalous Hall conductivity using the Fermi-loop approach
of Eq.~(\ref{eq:bp}) as applied to the three ferromagnetic transition
metals Fe, Co and Ni.

\subsection{bcc Fe}
\label{sec:fe}

We have previously presented calculations of the AHC of bcc Fe
computed using the Fermi-sea formulation.\cite{wang06} Here we
adopt the same choice of Wannier functions as in that work,
namely 18 Wannier functions covering the $s$, $p$ and $d$ character
and both spins.  The orbitals of $s$, $p$, and $e_g$ character are
actually rehybridized into Wannier functions of $sp^3d^2$ type, and
the Wannier functions are only approximate spin eigenstates because
of the presence of spin-orbit interaction (see Ref.~\onlinecite{wang06}
for details).

\begin{figure}
\begin{center}
   \epsfig{file=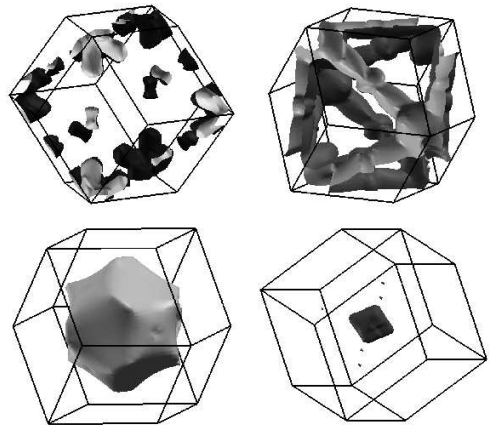,width=2.6in}
\end{center}
\caption{Calculated Fermi surfaces for bands 7-10 of bcc Fe (in order
of upper left, upper right, lower left, lower right). The outside 
frame is the boundary of the Brillouin zone.}
\label{fig:fs_fe}
\end{figure}

In our calculation for bcc Fe, bands 5-10 cross the Fermi energy.
Fig.~\ref{fig:fs_fe} shows the Fermi-surface sheets for bands 7-10,
plotted using the {\tt Xcrysden} package.\cite{xcrysden}   (Bands 5
and 6 give rise to small hole pockets, not shown.)  Some of these
sheets (especially 7 and 8) are
quite complicated but, as expected, they all conform to the
lattice symmetries. What is not clearly visible in these
plots are the tiny spin-orbit-induced splittings, which change the 
connectivity of the Fermi surface. As mentioned earlier, such features
play an important role in the AHC, and need to be treated with care.

We take the magnetization to lie along the $[001]$ axis.  Choosing
$\b_1=(2\pi/a)(1\bar 10)$ and $\b_2=(2\pi/a)(110)$ in the notation
of Sec.~\ref{sec:floop}, it follows that
$\a_3=2\pi\,\b_1\times \b_2/V_{\rm recip}=(0,0,a)$
where $V_{\rm recip}$ is the primitive reciprocal cell volume,
and we only need to compute the $c_{n3}$ in Eq.~(\ref{eq:Cshort}).
The slices are square in shape, and $k_\perp=k_z$ is
discretized into 500 slices.

\begin{figure}
\begin{center}
\epsfig{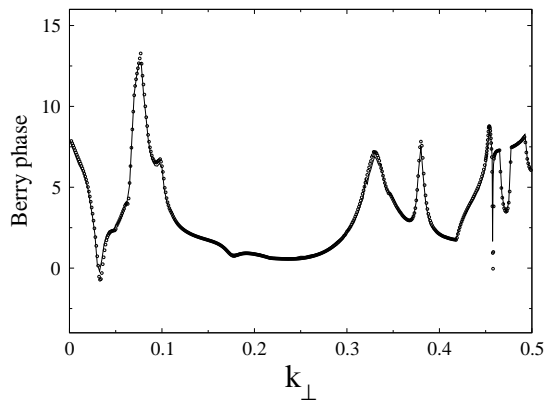}
\end{center}
\caption{Calculated Berry phase $\phi(k_\perp)$ of bcc Fe (in
radians) as a function of $k_\perp$ (in units of $2\pi/a$).  Solid
line shows results obtained from the Fermi-loop method of
Eq.~(\ref{eq:bp}); circles indicate reference results obtained by
the integration of the Berry curvature on each slice using
Eq.~(\ref{eq:phibc}).}
\label{fig:bp_fe}
\end{figure}

In Fig.~\ref{fig:bp_fe} we have plotted the total Berry phase
Eq.~(\ref{eq:phitot}) on each slice as computed from
Eq.~(\ref{eq:bp}).  The results are symmetric under mirror
symmetry, so only half of the range of $k_\perp$ is shown.  The
sharp peaks and valleys in Fig.~\ref{fig:bp_fe} are related to
degenerate or near-degenerate bands that have been split by the
spin-orbit interaction, as was illustrated, e.g., in the inset of
Fig.~\ref{fig:slice}.  To validate the calculation, we compare it
against a direct numerical integration of the Berry curvature over
the occupied bands using Eq.~(\ref{eq:phibc}), as indicated by the
symbols in Fig.~\ref{fig:bp_fe}.  In spite of rather complex and
irregular Fermi surfaces, the agreement between the two methods in
Fig.~\ref{fig:bp_fe} is excellent.

The values of the integrated anomalous Hall conductivity using
the new approach and the reference approach are shown in the first
and third lines of Table~\ref{table:ahc}.  The second line shows
the contribution obtained from integrating only the first term of
Eq.~(\ref{eq:bp}); clearly, this contribution is very small.
The agreement with the previous theory of Yao {\it et al.}\cite{yao04}
is excellent, while the agreement with experiment is only fair.
Table~\ref{table:ahc} will be discussed further in
Sec.~\ref{sec:discresults}.

Our approach also opens the possibility of discussing which Fermi
sheets are responsible for features visible in
Fig.~\ref{fig:bp_fe}.  For example, the dip near $k_\perp=0.03$ and
the peak near $k_\perp=0.33$ (in units of $2\pi/a$) come from
sheets 8-9, the peak near $k_\perp=0.09$ comes from sheets 6-8, and
the complex structure in the range of $k_\perp$ from 0.36-0.50
comes mainly from sheets 7-9.  Overall, the contribution from bands
5 and 10 are almost negligible, and bands 7-9 give much the largest
contributions.

\begin{table}
\caption{Anomalous Hall conductivity, in S/cm.
First three rows show values computed using
Eqs.~(\ref{eq:sigmaK})--(\ref{eq:Cshort}) together with
Eq.~(\ref{eq:bp}), the first term only of Eq.~(\ref{eq:bp}), or
Eq.~(\ref{eq:phibc}), respectively.  Results of previous theory and
experiment are included for comparison.}
\begin{ruledtabular}
\begin{tabular}{lrrr}
& bcc Fe & fcc Ni & hcp Co \cr
\hline
Fermi loop & 750\,\, & $-$2275\,\, & 478\,\, \cr
Fermi loop (1st term) & 7\,\, & 0\,\,  & $-$4\,\, \cr
Berry curvature & 753\,\, & $-$2203\,\, & 477\,\, \cr
Previous theory & 751\footnotemark[1]
    & $-$2073\footnotemark[2] & 492\footnotemark[2] \cr
Experiment& 1032\footnotemark[3]  & $-$646\footnotemark[4] &480\footnotemark[5]  \cr
\end{tabular}
\end{ruledtabular}
\footnotetext[1]{Ref.~\onlinecite{yao04}.}
\footnotetext[2]{Y. Yao, private communication.}
\footnotetext[3]{Ref.~\onlinecite{dheer67}.}
\footnotetext[4]{Ref.~\onlinecite{lavine61}.}
\footnotetext[5]{Ref.~\onlinecite{miyasato}.}
\label{table:ahc}
\end{table}

\subsection{fcc Ni}
\label{sec:ni}

For fcc Ni we chose 14 Wannier functions, seven each of approximately
spin-up and spin-down character.  These were comprised of five Wannier
functions of $d$-like symmetry centered on the Ni atoms and two Wannier
functions of tetrahedral symmetry located on the tetrahedral interstitial
sites, similar to the choice that was made for Cu in
Ref.~\onlinecite{souza01}.  The inner energy window was chosen to
extend 21\,eV above the bottom of the bands, thus extending  
$7.1$ eV  above the Fermi energy and including several
unoccupied bands as well.

\begin{figure}
\begin{center}
   \epsfig{file=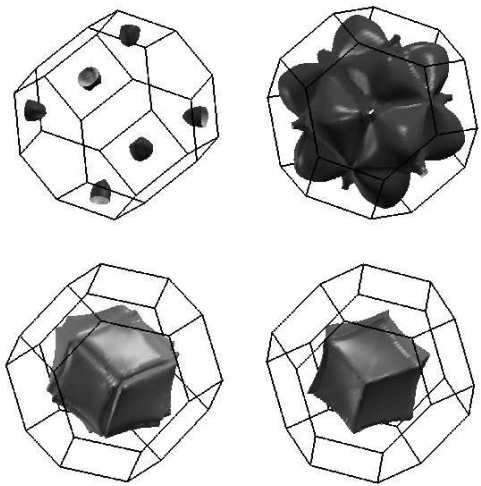,width=2.6in}
\end{center}
\caption{Calculated Fermi surfaces for bands 9-12 of fcc Ni. The outside 
frame is the boundary of the Brillouin zone.}
\label{fig:fs_ni}
\end{figure}

Our calculation is consistent with previous DFT calculations in
predicting that five bands (bands 8-12) cross the Fermi energy in
fcc Ni.  The Fermi sheets for bands 9-12 are shown in
Fig.~\ref{fig:fs_ni}.  Band 8 only barely crosses the Fermi energy
and gives rise to very small hole pockets near the X points (even
smaller than those illustrated for band 9).  The existence of
these pockets is a delicate feature that is not clearly confirmed
experimentally and is inconsistent with some recent LDA~+~$U$
calculations.\cite{yang01} However, including them or not has
very little influence on our calculated AHC, as explained below.
The shapes of the Fermi sheets in fcc Ni are somewhat more spherical
than those of bcc Fe. As expected, they
again conform to the lattice symmetries. 
 
In the case of fcc Ni, the magnetization lies along the $[111]$
axis.  Choosing $\b_1=(2\pi/a)(0\bar22)$ and
$\b_2=(2\pi/a)(20\bar2)$ in the notation of Sec.~\ref{sec:floop},
it follows that $\a_3=2\pi\,\b_1\times \b_2/V_{\rm
recip}=(a,a,a)=a\sqrt{3}\,\hat{\bf e}_{(111)}$, and we only need to
compute the $c_{n3}$ in Eq.~(\ref{eq:Cshort}).  The slices are
hexagonal in shape, and $k_\perp=\k\cdot\hat{\bf e}_{(111)}$ is
discretized into about $100$ slices.

\begin{figure}[b]
\begin{center}
\epsfig{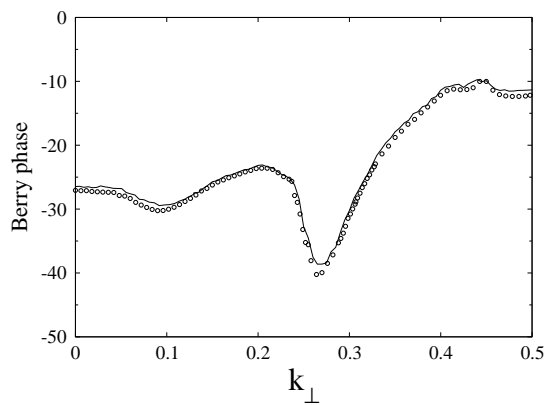}
\end{center}
\caption{Calculated Berry phase $\phi(k_\perp)$ of fcc Ni (in
radians) as a function of $k_\perp$ (in units of $2\pi/\sqrt{3}a$).
Solid line shows results obtained from the Fermi-loop method of
Eq.~(\ref{eq:bp}); circles indicate reference results obtained by
the integration of the Berry curvature on each slice using
Eq.~(\ref{eq:phibc}).}
\label{fig:bp_ni}
\end{figure}

The results are plotted in Fig. \ref{fig:bp_ni}, along with symbols
denoting the reference calculation by an integration of the Berry curvature over the slice. 
Once again, the agreement is very satisfactory.  The values
of the integrated AHC are again summarized in Table~\ref{table:ahc}.
A band-by-band analysis indicates that band 8 gives only a very
small contribution, less than 5\% in magnitude and opposite in sign,
to the total AHC. The hole pockets in band 9 give a
slightly larger positive contribution, but we find that the
dominant negative contribution to the AHC comes from bands 10-12.
 
\subsection{hcp Co}
\label{sec:co}

Co in the hcp structure has two atoms per unit cell. We choose 18 
Wannier functions per Co atom, nine for each spin, in a very similar
manner as was done for Fe in Sec.~\ref{sec:fe}.  We therefore have
36 Wannier functions per cell.

\begin{figure}
\begin{center}
   \epsfig{file=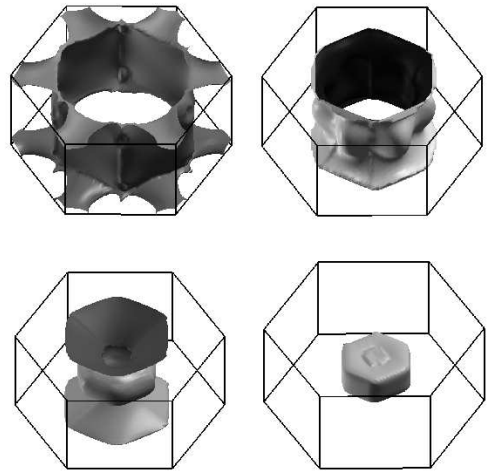,width=2.6in}
\end{center}
\caption{Calculated Fermi surfaces for bands 18-21 of hcp Co. The outside 
frame is the boundary of the Brillouin zone.}
\label{fig:fs_co}
\end{figure}

In our calculation, seven bands (bands 16-22)
cross the Fermi energy in hcp Co.  We show the four
largest Fermi-surface sheets associated with bands 18-21 in
Fig.~\ref{fig:fs_co}.  The Fermi
surfaces can be seen to respect the 6-fold crystal symmetry,
and none of them touch each other. 

The magnetization of hcp Co lies along the $[001]$ axis.
We thus choose $\b_1=(2\pi/a)(1/\sqrt{3},-1,0)$ and
$\b_2=(2\pi/a)(1/\sqrt{3},1,0)$ in the notation of Sec.~\ref{sec:floop},
and it follows that $\a_3=2\pi\,\b_1\times \b_2/V_{\rm
recip}=(0,0,c)$.  The slices are hexagonal in shape, and
$k_\perp=k_z$ is discretized into about $200$ slices.

\begin{figure}
\begin{center}
\epsfig{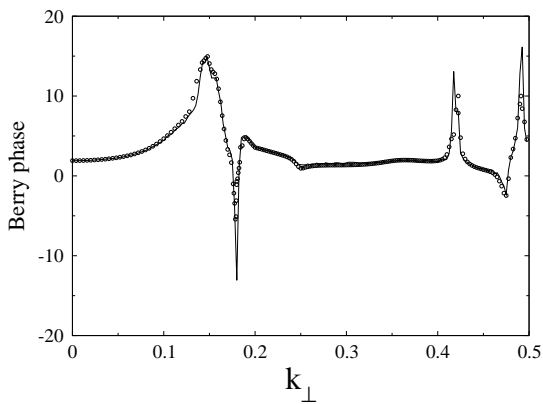}
\end{center}
\caption{Calculated Berry phase $\phi(k_\perp)$ of hcp Co (in
radians) as a function of $k_\perp$ (in units of $2\pi/c$).  Solid
line shows results obtained from the Fermi-loop method of
Eq.~(\ref{eq:bp}); circles indicate reference results obtained by
the integration of the Berry curvature on each slice using
Eq.~(\ref{eq:phibc}).}
\label{fig:bp_co}
\end{figure}

The results are plotted in Fig.~\ref{fig:bp_co}, along with the
symbols denoting the reference calculation by integration of the
Berry curvature.  Once again, the peaks and valleys correspond
to the places where two loops approach one another closely.
Some pieces of the Fermi surfaces of hcp Co are nearly parallel to
the slices (see the bottom right panel of Fig.~\ref{fig:fs_co}),
so that the number and shapes of the Fermi loops sometimes change
rapidly from one slice to another.  In particular, we found it
difficult to enforce continuity of the branch choice of
Eq.~(\ref{eq:bp}) as a function of $k_\perp$ near the sharp
features at $k_\perp a/2\pi=0.18$ and $0.42$ in
Fig.~\ref{fig:bp_co}.  We therefore redetermined the correct
branch choice by comparing with the result of the Berry-curvature
integration at slices just outside these difficult regions.
Despite these difficulties, it can still be seen that the
Fermi-loop method works well for this case.
Some of the sharp structure appearing in Fig.~\ref{fig:bp_co} in the
range of $k_\perp$ from 0.4-0.5 (in units of $2\pi/a$) arises from
the small hole pocket in band 16, but this gives a rather small
contribution to the total AHC.  The peak around $k_\perp=0.14$ and
the sharp dip around 0.18 comes mainly from the sheets associated
with bands 20 and 21.

\subsection{Discussion}
\label{sec:discresults}

\subsubsection{Internal consistency of the theory}

The second row of Table~\ref{table:ahc} shows the results computed
using only the first term of Eq.~(\ref{eq:bp}).
In each case, its contribution is
less than 1\% of the total, and would therefore be negligible for
most purposes.
Actually, it can be shown that the inclusion of the first term only
in Eq.~(\ref{eq:bp}) of the present method is equivalent to
carrying out the Berry-curvature integration approach of
Ref.~\onlinecite{wang06} with the $D$--$D$ term omitted in Eq.~(32)
of that work (that is, only the $D$--$\overline \rm A$ and
$\overline\Omega$ terms included).  We have carried out this comparison
and find values of 7, $-$0.5 and $-$2\,S/cm for
bcc Fe, fcc Ni, and hcp Co, respectively, in very good agreement
with the values reported in Table~\ref{table:ahc}.
The physical interpretation for the small terms in the second row
of Table~\ref{table:ahc} is basically that
the full set of Bloch-like states constructed from the Wannier
functions (e.g., the manifold of 18 Bloch-like states in bcc Fe)
has some small Berry curvature of its own, and the projection of
this curvature onto the occupied subspace gives the small first
term of Eq.~(\ref{eq:bp}).  On the other hand, spin-orbit induced
splittings across the Fermi level {\it between} Bloch-like 
states built from these
Wannier functions give large, sharply peaked contributions to the
Berry curvature of the occupied subspace,
and make a very much larger contribution to the
total AHC.  Of course, the precise decomposition between the first
and second term of Eq.~(\ref{eq:bp}) depends on the exact choice
of Wannier functions, but the present results seem to indicate that
the dominance of the second term is probably a general feature,
at least for systems in which the Wannier functions are well
localized and the spin-orbit splitting is not very strong.

As mentioned in the previous section, the overall agreement 
seen in Table~\ref{table:ahc} between the
results computed using the Fermi-loop approach and those computed
using the Berry-curvature integration indicate the internal consistency of our
theory and 
implementation.  The agreement with the results of Yao and coworkers,
which were obtained by a Berry-curvature integration using an
all-electron approach,\cite{yao04} also demonstrates the robustness
of our pseudopotential implementation, including its ability to
represent spin-orbit interactions correctly.

\subsubsection{Comparison with experiment}

In the last row of Table~\ref{table:ahc} we show comparison with
some representative experimental values for the AHC of Fe, Ni, and Co.  
However, it should be kept in mind that there is some uncertainty
and variation in the values reported by different groups.  For example,
Ref.~\onlinecite{smit55} gives a value for Ni 
of $-$753\,S/cm and Ref.~\onlinecite{Jellinghaus61} 
reports a value for Co of 500\,S/cm.  It could well be 
that different kinds of experimental samples have different impurity and defect
populations, leading to different extrinsic contributions to
the AHC.  Since the theoretical values presented in Table~\ref{table:ahc}
are all computed by including only the intrinsic Karplus-Luttinger
contribution to the AHC, so that extrinsic skew and side-jump scattering
contributions are neglected, it is most appropriate to compare with
experimental measurements in which the effects of the intrinsic
contribution are isolated.

A serious effort in this direction has recently been made
by studying and correlating the variation of both the longitudinal and
the anomalous Hall conductivity as a function of temperature.\cite{miyasato}
It was found that for Fe, Co, and Ni, $|\sigma_{xy}|$ remains roughly constant
between 150 and 300\,K while $\sigma_{xx}$ changes by about a factor
of four.
The value of $\sigma_{xy}$ in this plateau was attributed to the
intrinsic mechanism, which should be independent
of the scattering rate. The values thus obtained are
about 970 and $-$480\,S/cm for Fe and Ni films
respectively (the
value quoted in Table~\ref{table:ahc} for Co from the same work is also a 
film value), and about 2000\,S/cm for single-crystal Fe.
The factor-of-two difference reported in Ref.~\onlinecite{miyasato}
between the intrinsic $\sigma_{xy}$ of Fe in the single-crystal and
film forms is puzzling and deserves further investigation.

Turning now to the comparison between theory and experiment, we
find a very rough agreement at the level of signs and general
trends. However, the agreement is not quantitatively accurate,
except for Co where the agreement is good.  For Fe our
results are in very rough agreement ($\sim$25\% low) compared to
the results of Ref.~\onlinecite{dheer67} or the film results of
Ref.~\onlinecite{miyasato}, but a factor of two smaller than the
single-crystal results of Ref.~\onlinecite{miyasato}.  Clearly the
most serious discrepancy is for Ni, for which we get a consistent
sign but a much larger magnitude than indicated by the
experiments.  (The issue\cite{yang01} of whether a hole pocket
really appears in band 8 of fcc Ni is not relevant since, as indicated
in Sec.~\ref{sec:ni}, it makes a quite small numerical contribution
to our theoretical result.)  While the available experimental
values for Ni appear to be roughly consistent with each other, we
are not aware of any study using the methods of
Ref.~\onlinecite{miyasato} applied to both films and single
crystals of Ni.  Until the experimental situation is clarified
further, a final judgment on the degree of disagreement with the
values based on DFT calculations should perhaps be withheld.

In the meantime, it would be desirable to face some of the
challenges and open questions that remain on the theoretical side.
For example, not much is yet known about the accuracy of common
exchange-correlation functionals, such as the PBE functional used
here,\cite{perdew96} for computing the AHC.  Fe, Ni and Co have open
$d$ shells and can be considered from one point of view to be
strongly-correlated systems.  The fact that the magnetic moments are
given accurately by DFT (see Table \ref{table:spin_moment}) does
not necessarily mean that more delicate properties, especially those
like the AHC and the magnetocrystalline anisotropy\cite{yang01}
that depend on spin-orbit interactions, will be given accurately.
It is possible that the use of more sophisticated density-functional
approaches (e.g., current-density functional theories) or higher-level
many-body approaches may ultimately prove necessary.  Finally, it
would be desirable to develop DFT-based methods for computing the
defect-related extrinsic contributions, but this will also prove to
be a daunting challenge, not least because the relevant defect
populations are not known.

In summary, experiments and DFT-based theories agree on the
orders of magnitude and signs of the intrinsic Karplus-Luttinger
contributions to the AHC in these three ferromagnetic metals,
and the results for Fe and especially Co suggest that quantitative
agreement may be obtainable.  The substantial discrepancy for
Ni deserves further attention.

\section{Computational efficiency}
\label{sec:disc}

The motivation for developing a method for computing the AHC that
relies only on information computed on the Fermi surface is, to some
degree, esthetic and philosophical: Haldane argued that the
AHC is physically most naturally regarded as a Fermi-surface
property,\cite{haldane04} and as such should be computed using a
method that does not make use of extraneous information in arriving
at the desired quantity.  However, a much more important motivation
from the practical point of view is the idea that the computational
effort might be drastically reduced by having to compute
quantities only on the two-dimensional Fermi surface rather than on a
three-dimensional mesh of $k$-points.

In the present implementation as it stands, unfortunately, the
computational savings gained through the use of the Fermi-loop
Berry-phase approach is quite modest.  After taking advantage of
the symmetry as discussed in Sec.~\ref{sec:sym}, the total
computational time of our AHC calculation for bcc Fe is about 1.7
hours using a $200\times 200$ k-mesh on each of 500 slices, to be
compared with about 2 hours using our previous method of
Ref.~\onlinecite{wang06}.  (These timings are on a 2.2 GHz
AMD-Opteron PC, and neither includes the Wannier construction
step, which takes about 2.5 hours.)  Roughly, the work on each slice can be divided into three
phases: Step 1, computing the energy eigenvalues on the $200\times
200$ k-mesh; Step 2, executing the contour-finding algorithm; and
Step 3, evaluating Eq.~(\ref{eq:bp}) on the discretized Fermi loops.
We find that less than 1\% of the computer time goes to Step 2,
while the remainder is roughly equally split between Step 1 and
3.  The operations in these steps have been greatly accelerated
by making use of Wannier interpolation methods, but this is also the
case for the comparison method of Ref.~\onlinecite{wang06}.
(We emphasize that, for this reason, both the method of
Ref.~\onlinecite{wang06} and the present one are orders of
magnitude faster than methods based on direct first-principles
calculations at every $k$-point.)

Many opportunities for further reduction of the computer time
are worthy of further exploration.  Regarding Step 1, for example,
at the moment the contour-finding is done independently
on each slice; it might be much more efficient to step from slice
to slice and use a local algorithm to determine the
deformation of the Fermi contours on each step.  It may also be
possible to do a first cut at the contour-finding using a coarser
k-mesh (say $50\times50$) and then refine it in regions where the
loops approach one another or have sharp bends.  It may also be
possible to take larger steps between slices in most regions of
$k_\perp$, and fall back to fine slices only in delicate regions.
In implementing all such strategies, however, one should be careful
to avoid missing any small loops that might appear suddenly from
one slice to the next, or which might be missed on an initial
coarse sampling of the slice.   It may also be interesting to
explore truly three-dimensional algorithms for finding contour
surfaces, and then derive two-dimensional loops from these.

As for Step 3, it should be possible to use a lower density of
$k$-points in the portions of the loop discretization where the
character of the wavefunctions is changing slowly.  The time for
this step will also obviously benefit from taking larger steps
between slices in regions where this is possible.  Finally, a
reduction by a factor of two or more may be possible by making
use of symmetries not considered in Sec.~\ref{sec:sym}, such as
the diagonal mirror symmetries ($x\leftrightarrow y$ etc.) in
bcc Fe.

The exploration of these issues is somewhat independent from the
quantum-mechanical formulation of the underlying theory, which is
the main focus of the present work, and we have therefore left the
exploration of these possibilities for future investigations.

Finally, it should be emphasized that the computational load scales
strongly with the dimension of the Wannier space used to represent
the wavefunctions.  In our calculations, this was 18, 14, and 36 for
Fe, Ni, and Co, respectively.  In some materials, there may be only
a few bands crossing the Fermi energy, and it might be possible to
represent them using a much smaller number of Wannier functions.
This is the case in many transition-metal oxides such as
Sr$_2$RuO$_4$, cuprate superconductors, etc.  In ferromagnetic
materials of this kind, it should be possible to choose an inner
window in the Wannier disentanglement procedure\cite{souza01}
that brackets the Fermi energy but does not extend to the bottom
of the occupied valence band, and to generate just a handful
of Wannier functions (e.g., three $t_{2g}$ orbitals times two for
spin) to be used in the Wannier interpolation procedure.  Then all
matrices used in that procedure would be very much smaller (e.g.,
6$\times$6) and the computation would go considerably faster.

\section{Summary}
\label{sec:summary}

In summary, we have developed a first-principles method for 
computing the intrinsic AHC of ferromagnets as a Fermi-surface
property.  Unlike conventional methods that are based on a $k$-space
volume integration of the Berry curvature over the occupied Fermi sea,
our method implements the Fermi-surface philosophy by dividing the
Brillouin zone into slices normal to the magnetization direction
and computing the Berry phases of the Fermi loops on these slices.
While Haldane has pointed out that only the non-quantized part of
the AHC can be determined in principle from a knowledge of
Fermi-surface properties only, we find in practice that it is
straightforward to make the correct branch choice and resolve the
quantum of uncertainty by doing a two-dimensional Berry-curvature
integration on just one or a few of the slices.  Our method also
makes use of methods of Wannier interpolation to minimize the
number of calculations that have to be done using a full first-principles
implementation; almost all the operations needed to compute the
AHC are actually done by working with small matrices (e.g.,
18$\times$18 for bcc Fe) in the Wannier representation.  The
new method also allows us to discuss the contributions to the
AHC arising from individual Fermi sheets or groups of sheets.

We have tested and validated our new method by comparing with our
earlier implementation of a Fermi-sea Berry-curvature integration
for bcc Fe, fcc Ni, and hcp Co.  The different crystal structures
and magnetization orientations in these three materials also allow
us to demonstrate the flexibility of the method in dealing with
these different cases.  We find excellent agreement between the two
approaches in all cases.

\acknowledgments

This work was supported by NSF Grant DMR-0549198.  We wish to thank
Mark Stiles for useful discussions.



\end{document}